# Particle Swarm Optimization for Great Enhancement in Semi-Supervised Retinal Vessel Segmentation with Generative Adversarial Networks

Qiang Huo[1], Geyu Tang[1], and Feng Zhang[1]

[1] Institute of Microelectronics of Chinese Academy of Sciences, Beijing 100029, China, and with the University of Chinese Academy of Sciences, Beijing 100049, China.
zhangfeng_ime@ime.ac.cn

**Abstract.** Retinal vessel segmentation based on deep learning requires a lot of manual labeled data. That's time-consuming, laborious and professional. In this paper, we propose a data-efficient semi-supervised learning framework, which effectively combines the existing deep learning network with generative adversarial networks (GANs) and self-training ideas. In view of the difficulty of tuning hyper-parameters of semi-supervised learning, we propose a method for hyper-parameters selection based on particle swarm optimization (PSO) algorithm. This work is the first demonstration that combines intelligent optimization with semi-supervised learning for achieving the best performance. Under the collaboration of adversarial learning, self-training and PSO, we obtain the performance of retinal vessel segmentation approximate to or even better than representative supervised learning using only one tenth of the labeled data from DRIVE.

## 1    Introduction

Retinal vessel segmentation is very important for assistant diagnosis, treatment and surgical planning of fundus diseases. Retinal vessel segmentation is also a necessary step for accurate visualization and quantification of retinal diseases. Changes of vascular morphology, such as shape, tortuosity, branch and width, provide a powerful reference to ophthalmologists in early diagnosis of retinal diseases [1]. Automatic vessel segmentation is of great value to improve work efficiency and reduce errors caused by fatigue and non-uniform illumination.

Conventional automatic retinal vessel segmentation includes supervised and unsupervised methods. Usually, most of unsupervised methods are rule-based technologies, including traditional matched filtering [2], morphological processing [3], vessel tracing [4], thresholding [5], etc. Although unsupervised learning usually has the advantage of high speed and low computational complexity, it requires hand-crafted feature extraction, which needs professional knowledge and is difficult to achieve high accuracy. Supervised methods are to train a series of gold standard data for pixel-level classification. Gold standard images are usually labeled by experienced ophthalmologists [6]. Supervised methods are widely used because of its better performance than unsupervised method without tricky manual feature engineering. However, little work about retinal vessel segmentation has been done under the semi-supervised learning framework.

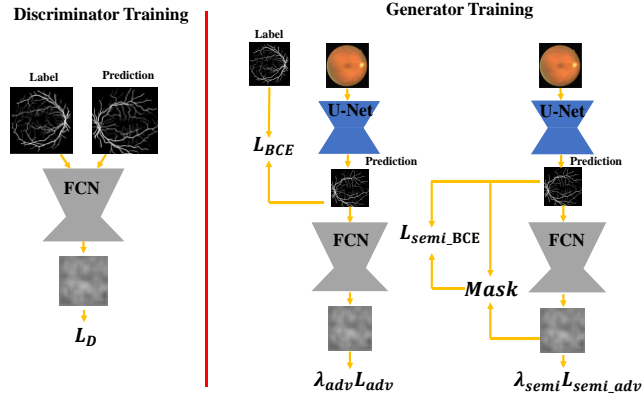

**Fig. 1.** Visualization of training of discriminator and segmentation networks for proposed semi-supervised learning.

In recent years, convolutional neural network (CNN) has been successfully applied in computer vision. Some studies have also extended it to the field of retinal vessel segmentation and achieved similar or even more than human segmentation results. However, these methods require huge number of artificial labeled images. Fundus images are not only difficult to obtain, but also require enormous efforts from ophthalmologists. There are obstacles in the wide application of supervised retinal vessel segmentation based on CNN [7-8]. GAN has excellent ability to learn potential distribution and has been successfully used in unsupervised and semi-supervised learning [9]. GAN consists of generator and discriminator. Generators try to generate distributions close to the original data, while discriminators try to distinguish the original data from the generated data [10].

Particle swarm optimization (PSO) algorithm is an intelligent optimization algorithm based on the idea that the social sharing of information among the same species can bring benefits. Because of its concision, easy implementation, no need gradient information and few parameters, PSO has shown good results in continuous optimization problems and discrete optimization problems, especially its natural real coding characteristics are suitable for solving the problems of real optimization [11]. PSO has been applied to the selection of hyper parameters of deep neural networks and achieved excellent results [12].

In this paper, we replace the original generation network of GAN with segmentation network. The discriminator network uses the fully convolutional neural network with the confidence map as output instead of the discriminator which only outputs a single probability value. We introduce self-training scheme to construct semi-supervised learning framework with GAN for unlabeled data. PSO is proposed to select semi-supervised learning parameters to avoid its own noise. The scheme can greatly improve the effect of semi-supervised learning, save a lot of time and reduce the technical threshold compared with manual tuning.

## 2 Proposed methods

### 2.1 Network Architecture

U-Net has the structure that initial convolutional feature maps are skip-connected to up-sampling layers from bottleneck layers. The structure can transfer the low-level features such as edges and lines of the original feature maps to the up-sampling layers, which makes the segmentation network better recognize the details, especially in the medical field with little data [13]. Therefore, we adopt U-Net as our segmentation network in our work, which is given a fundus image labeled or unlabeled and generates a probability map with the same size as input. The probability map shows the probability that each pixel belongs to the vessel.

We adopt the simplified fully convolutional network as our discriminator network. It consists of 5 convolution layers with $4 \times 4$ kernel and {64,128,256,512,1} channels in the stride of 2 followed by a Leaky-ReLU parameterized by 0.2 except the last layer and an up-sampling layer to get the probability map with the size of the input map. Figure 1 shows our proposed algorithm framework.

### 2.2 Loss Function

The fundus images can be recognized as $X_n$ with three channels. The output probability map of segmentation network can be regarded as function $S(X_n)$. $S(X_n)^{(h,w)}$ indicates the probability value of each pixel derived from the vessel. The discriminator network $D(\cdot)$ takes the output probability map $S(X_n)$ and the ground truth label map $Y_n$ as input.

The loss function $L_{SEG}$ of our segmentation network can be defined as,

$$L_{SEG} = L_{BCE} + \lambda_{adv}L_{adv} + \lambda_{semi\_adv}L_{semi\_adv} + \lambda_{semi\_BCE}L_{semi\_BCE} \quad (1)$$

where,

$$L_{BCE} = -\frac{1}{nhw}\sum_{n,h,w} y_n^{(h,w)} log\big(S(X_n)^{(h,w)}\big) + (1 - y_n^{(h,w)})log(1 - S(X_n)^{(h,w)}) \quad (2)$$

$$L_{adv} = -\frac{1}{nhw}\sum_{n,h,w} log\left(D\big(S(X_n)\big)^{(h,w)}\right) \quad (3)$$

$$L_{semi\_BCE} = -\frac{1}{nhw}\sum_{n,h,w} I\left(D\big(S(X_n)\big)^{(h,w)} > T_{semi_{mask}}\right)(\hat{y}_n^{(h,w)} log\big(S(X_n)^{(h,w)}\big) + \\ (1 - \hat{y}_n^{(h,w)})log(1 - S(X_n)^{(h,w)})) \quad (4)$$

The components of the loss function are:

    i.   $L_{BCE}$ is the usual binary cross entropy loss of labeled fundus images. The target of this is to maximize the predicted probability over the correct class label.

ii. $L_{adv}$ and $L_{semi\_adv}$ are the adversarial losses of predicted probability map generated from labeled and unlabeled fundus images by segmentation network. They have the same expression except for the different input sources. $\lambda_{adv}$ and $\lambda_{semi\_adv}$ are coefficients of $L_{adv}$ and $L_{semi\_adv}$ respectively.

iii. $L_{semi\_BCE}$ are the binary cross entropy loss of unlabeled fundus images based on a self-training semi-supervised learning framework. As shown in the indicator function $I(D(S(X_n))^{(h,w)} > T_{semi\_mask})$, the main idea of self-training is that the discriminator estimates the probability map of unlabeled data and selects the most assured pixels to participate in the construction of loss function. Then the label of selected pixels with high confidence is element-wise set with $\hat{y}_n^{(h,w)} = 1$ if $S(X_n)^{(h,w)} > 0.5$, and $\hat{y}_n^{(h,w)} = 0$ otherwise. $\lambda_{semi\_BCE}$ is the coefficient of $L_{semi\_BCE}$.

The loss function $L_D$ of our discriminator can be defined as,

$$L_D = -\frac{1}{nhw} \sum_{n,h,w} log\left(1 - D\big(S(X_n)\big)^{(h,w)}\right) + \log\big(D(Y_n)^{(h,w)}\big) \qquad (5)$$

where $Y_n$ represents the vessel ground truth label map of fundus image. The adversarial loss is similar with traditional GAN in which target is to enhance the ability of discriminator to distinguish positive samples or others.

### 2.3 Hyper-Parameter Selection for Semi-supervised Learning

Since self-training based semi-supervised learning is sensitive to hyper-parameters, it is easy to recognize unlabeled data as noise without refined parameter options. However, manual tuning will take a lot of time and be confusing. We deploy PSO as a wrapper to excavate hyper-parameters that make full use of unannotated fundus images, including $\lambda_{semi\_adv}$, $\lambda_{semi\_BCE}$ and $T_{semi\_BCE}$. Algorithm 1 shows the complete step of hyper-parameter selection for semi-supervised learning.

## 3 Implementation Details

### 3.1 Evaluation datasets and metric

We conduct experiments on DRIVE datasets [14]. DRIVE dataset has a clear demarcation of training and test set with 20 images in each category. The first manual annotator's images are used to train and test our proposed model. During training, we use the random scaling and cropping operations with size 512×512. We train all our models on this dataset for 20K iterations with batch size 2. We evaluate our methods with Area Under Curve for Receiver Operating Characteristic (ROC AUC), Area Under Curve for Precision and Recall curve (PR AUC) and the mean of them (Score).

### 3.2 Training details

Our methods are implemented based on PyTorch library. We train the proposed model on a single GTX 1080Ti GPU with 11GB memory. To train our U-Net, we use the Adam optimizer with fixed learning rate of 1e-4 and $\beta_1 = 0.5$, $\beta_2 = 0.9$. For hyper-

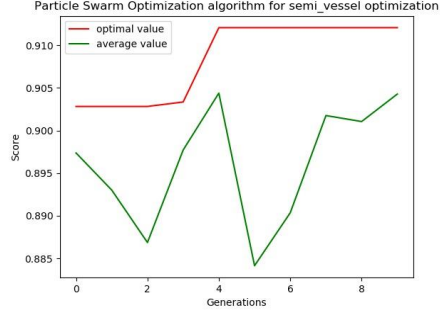

**Fig. 2.** The results of PSO for selecting the semi-supervised learning hyper-parameters when using 0.5 labeled images.

---

Algorithm 1 PSO for selecting the hyper-parameters of semi-supervised learning.

---

**Input:** $b_l, b_u, G, S, \phi_p, \phi_g, \omega$            **Output:** $\hat{P}$.

begin

         $f(P^S) \leftarrow -\infty$

for each particle $i = 1, \dots, S$ do

     Initialize the particle's position with a uniformly        distributed random vector: $P_i \sim U(b_l, b_u)$

     Initialize the particle's best position to its initial position: $P_i^* \leftarrow P_i$

     if $f(P_i^*) > f(P^S)$ then

         Update the swarm's best position: $P^S \leftarrow P_i^*$

     Initialize the particle's velocity: $v_i \sim U(-|b_u - b_l|, |b_u - b_l|)$

while $g \leq G$ do

     for each particle $i = 1, \dots, S$ do

         Pick random numbers: $r_p, r_g \sim U(0,1)$

         Update the particle's velocity: $v_i \leftarrow \omega v_i + \phi_p r_p (P_i^* - P_i) + \phi_g r_g (P^S - P_i)$

         Update the particle's position: $P_i \leftarrow P_i + v_i$

         if $f(P_i) > f(P_i^*)$ then

             Update the particle's best position: $P_i^* \leftarrow P_i$

             if $f(P_i^*) > f(P^S)$ then

                 Update the swarm's best position: $P^S \leftarrow P_i^*$

                 $g \leftarrow g + 1$

         $\hat{P} \leftarrow P^S$

end

- - - - - - - - - - - - - - - - - - - - - - - - - - - - - - - - - - - - - - - - - -

$\hat{P} = (\lambda_{semi\_adv}, \lambda_{semi\_BCE}, T_{semi\_BCE})$: outputs vector of hyper-parameters for semi-supervised learning.

$b_l, b_u$: the lower and upper limits. $G$: number of iterations.

$S$: population size.      $\phi_p, \phi_g$: the acceleration coefficients.

$\omega$: inertia weight.

---

parameters in $L_{SEG}$, we set $\lambda_{adv}$ as 0.1. $\lambda_{semi\_adv}, \lambda_{semi\_BCE}$ and $T_{semi\_BCE}$ are determined by PSO. To train the discriminator, we also adopt the Adam optimizer with $\beta_1 = 0.5, \beta_2 = 0.9$. Unlike U-Net, the learning rate is fixed as 0.01e-4 to balance adversarial learning.

**Table 1.** Progressive comparison of our semi-supervised retinal vessel segmentation model on DRIVE dataset with respect to Area Under Curve (AUC) for Receiver Operating Characteristic (ROC), Precision and Recall (PR).

| Case I | | | |
|---|---|---|---|
| Model (Data Amount) | AUC (ROC) | AUC (PR) | Score |
| U-Net (0.1 labeled) | 0.9391 | 0.8051 | 0.8721 |
| U-Net+$L_{adv}$ (0.1 labeled) | 0.9496 | 0.8290 | 0.8893 |
| U-Net+$L_{adv}$+$L_{semi}$ (0.1 labeled+0.9 unlabeled) | 0.9550 | 0.8419 | 0.8985 |
| Case II | | | |
| Model (Data Amount) | AUC (ROC) | AUC (PR) | Score |
| U-Net (0.5 labeled) | 0.9554 | 0.8508 | 0.9031 |
| U-Net+$L_{adv}$ (0.5 labeled) | 0.9650 | 0.8547 | 0.9099 |
| U-Net+$L_{adv}$+$L_{semi}$ (0.5 labeled+0.5 unlabeled) | 0.9681 | 0.8676 | 0.9179 |

**Table 2.** Comparison of our semi-supervised retinal vessel segmentation model with other classical methods on DRIVE datasets with respect to Area Under Curve (AUC) for Receiver Operating Characteristic (ROC), Precision and Recall (PR). Note that our model uses only 0.1 labeled images and 0.9 unlabeled images from DRIVE.

| Method | AUC (ROC) | AUC (PR) | Score |
|---|---|---|---|
| Kernel Boost [6] | 0.9306 | 0.8464 | 0.8885 |
| Wavelets [16] | 0.9436 | 0.8149 | 0.8793 |
| HED [15] | 0.9696 | 0.8773 | 0.9234 |
| DRIU [8] | 0.9793 | 0.9064 | 0.9428 |
| Ours | 0.9550 | 0.8419 | 0.8985 |

The discriminator and the U-Net are trained alternatively in every iteration until all of them converge. The typical maximum number of iterations is set as 20K. For our proposed semi-supervised learning, it is noted that $\lambda_{semi\_BCE}$ is set to 0 before 5000 iterations for the loss function $L_{SEG}$ of U-Net in order to eliminate the noise introduced by immature U-Net and discriminator. The noise can be expressed as unreliable $I(D(S(X_n))^{(h,w)} > T_{semi\_mask})$ and $\hat{y}_n^{(h,w)}$ in Eq.4. Besides, our proposed PSO method for hyper-parameters selection of semi-supervised learning can remove some noises further.

### 3.3 Hyper-parameter selection details

For hyper-parameters selection based on PSO, the main procedure is depicted in Algorithm 1. $b_l$ and $b_u$ determine the lower and upper bounds of vector P_i of hyper-parameters for semi-supervised learning respectively. According to prior knowledge of our work, we set $b_l = (0,0,0)$ and $b_u = (0.01, 0.3, 0.5)$. The number of total iterations $G$ and the population size $S$ of particle swarm are set to 10 and 3 after considering the trade off between time complexity and performance. The acceleration coefficients $\phi_p$ and $\phi_g$ are set as 1. Inertia weight $\omega$ plays an important role in the convergence of PSO. We use time-varying weight to avoid crossing the global optimum due to the narrow range of particle positions and too fast flight speed. $\omega_g$ can be expressed as,

$$\omega_g = \omega_{max} - \frac{\omega_{max} - \omega_{min}}{N_{iter}} \times g \qquad (6)$$

where g is the number of iteration, $\omega_{max}$ is set as 0.5 and $\omega_{min}$ is set as 0.1.

We randomly choose two images from train dataset used to evaluate our fitness function $f(\cdot)$, that is, the score of our proposed model on retinal vessel segmentation. To reduce time of optimization and ensure optimal effect, the number of total iterations of each particle is modified to 1000 from 20000. Reference output vectors of hyper-parameters for semi- supervised learning are available after 16 hours of optimizing and training on a single GTX 1080Ti GPU. The optimal and average scores for each generation of our proposed models are shown in Figure 2. In this work, the optimal $\hat{P}$ is about (0.004, 0.1, 0.1).

## 4      Experimental Results

### 4.1      Progressive comparison of our own models

We randomly select 0.1 and 0.5 images from DRIVE dataset respectively as labeled data and the others as unlabeled data for training our semi-supervised retinal vessel segmentation framework. Case I and case II in Table 1 shows the progressive comparison of our own model. For our baseline, the U-Net achieves lowest score trained on 0.1 or 0.5 labeled images from DRIVE dataset. If combining adversarial learning with U-Net, the score will increase slightly because adversarial loss helps U-Net to generate probability map approximate to labeled data in details. After adding our semi-supervised loss, performance will be enhanced further to verify the effect of our proposed methods.

### 4.2      Comparison of different models

The comparison between our semi-supervised retinal vessel segmentation model and other representational methods is depicted in Table 2. Although only 0.1 labeled data is used, we have achieved similar or even better results compared with supervised learning using all labeled data from DRIVE.

## 5      Conclusions

In this paper we introduce a semi-supervised retinal vessel segmentation framework based on the idea of self-training and GAN. We adopt U-Net as our segmentation network given both labeled and unlabeled fundus image. For labeled data, the same method is adopted as traditional supervised learning. Inspired by self-training, the high confidence pixels of unbaled data are selected by well-trained discriminator and then are marked as 0 or 1 to construct virtual label through the probability map generated from U-Net. Besides, PSO has been successfully applied to select the hyper-parameters of semi-supervised learning, which is beneficial for excavating maximum information from unlabeled dada. Under the collaboration of adversarial learning, self-training and PSO to select optimal hyper-parameters, we obtain the score of retinal vessel segmentation approximate to or even better than supervised learning using only 0.1 labeled data finally.